\documentclass[twoside,10pt]{article} 

\usepackage{indentfirst} 
\usepackage{bm} 
\usepackage{graphicx} 
\usepackage{amsmath} 

\begin{document} 


\begin{center}
\LARGE\bf {Decoherence effect on quantum correlation and entanglement in a two-qubit spin chain}
\end{center}
\begin{center}  
 Mohammad Reza Pourkarimi$^{1,2 \dagger }$. \ Majid Rahnama$^{1}$. Hossein Rooholamini$^{1}$
\end{center}
\begin{center}  
\begin{small} \sl
{1.Faculty of physics, Shahid Bahonar University of Kerman, 76169-14111, Kerman, Iran}

{2.Physics Department, Salman Farsi University of Kazerun, Kazerun, Iran}
\end{small}
\end{center}
\vspace*{2mm}

\begin{center}  
\begin{minipage}{15.5cm}
\parindent 20pt\small
Assuming a two-qubit system in Werner state which evolves in Heisenberg XY model with  Dzyaloshinskii-Moriya (DM) interaction under the effect of different environments. We evaluate and compare quantum entanglement, quantum and classical correlation  measures. It is shown that in the absence of decoherence effects,  there is a critical value of DM interaction for which  entanglement may vanish while quantum and classical correlations do not. In the presence of environment the behavior of correlations depends on the kind of system-environment interaction. Correlations can be sustained by manipulating Hamiltonian anisotropic-parameter in a dissipative environment. Quantum and classical correlations are more stable than entanglement generally.
\end{minipage}
\end{center}

\begin{center}  
\begin{minipage}{15.5cm}
\begin{minipage}[t]{2.3cm}{\bf Keywords:}\end{minipage}
Quantum entanglement. Quantum correlation. Quantum discord. Classical correlation. Isotropic parameter. DM interaction. Hamiltonian XY model. Environment.
\end{minipage}
\end{center}


\footnotetext[2]{Corresponding author. E-mail: mrpourkarimy@gmail.com  }

\section{Introduction}  
Correlation between a bipartite system is divided into two parts: quantum correlation and classical correlation. One kind of quantum correlation is quantum entanglement which is not separable and has no classical counterpart.  Entanglement is a fundamental feature of quantum mechanics \cite{label1,label2,label3} which has many  applications in quantum information processing \cite{label4}, such as quantum cryptography, quantum computation and quantum communications \cite{label5,label6,label7}. 

Despite many applications of entanglement in quantum information processing, it also has some shortcomings, such as entanglement sudden death (ESD). This phenomenon happens when entanglement between subsystems vanishes during evolution in finite time \cite{label8,label9}. However, entanglement does not constitute all of the quantum correlation possible. Quantum correlation contains non-classical correlations even for some separable states \cite{label10}.  This shows the advantage of using quantum correlation in quantum information processing.

Although there is no known relationship between quantum entanglement and quantum correlation measures, some analytical and numerical comparisons have been made between them for quantum states in recent years \cite{label11,label12}. There are also lots of studies on quantum entanglement and quantum correlation dynamics in open and closed quantum systems \cite{label13,label14,label15,label16,label17,label18,label19,label20}. 

It may be interesting to study the classical correlation besides quantum correlation and entanglement. Therefore in this study, we calculate and compare the time evolution of  quantum entanglement, quantum correlation  and classical correlation measures for a  two-qubit system in Werner state under the influence of XY model Hamiltonian with spin-orbit interaction known as Dzyaloshinskii-Moriya (DM) interaction and various system-environment couplings. We will show that, different correlations have different behaviors, however, at the same conditions, quantum and classical correlations are more stable than quantum entanglement.  

\section{ The model and its solution}

We assume the Werner state \cite{label21} as the initial state: 
\begin{equation}\label{initial state}
 \rho(0)=(\frac{1-p}{4}I_{4\times4}+p|\varphi><\varphi|),
\end{equation}
where $|\varphi>=\frac{1}{\sqrt{2}}(|10>+|01>)$ is the maximally entangled state of two qubits (i.e. the Bell state), $I_{4\times4}$ is an identity  $4\times4$ matrix and $p$ is the pureness degree of the state  in which  $\rho$ is not entangled for $0\leq p\leq\frac{1}{3}$ and entangled for $\frac{1}{3}< p\leq1$. We evaluate the dynamics of the initial state (\ref{initial state}) as follows: \cite{label22}

\begin{align}\label{master}
\frac{d\rho(t)}{dt}&=-i[H,\rho(t)]+\frac{\gamma}{2} \sum_{j=1,2}(2 b_{j}\rho(t) b^\dag_{j}- b^\dag_{j} b_{j} \rho(t)- \rho(t) b^\dag_{j} b_{j} )
\nonumber \\
&=-i[H,\rho(t)]+\ell_\rho,
\end{align}
where $H$  is the Hamiltonian for a two-qubit system in an anisotropic Heisenberg XY model with DM interaction defined as \cite{label23,label24}
\begin{equation}\label{Hxy1}
H=\frac{1}{2}( J_x \sigma^x_1 \sigma^x_2+J_y \sigma^y_1 \sigma^y_2  +D(\sigma^x_1 \sigma^y_2-\sigma^y_1 \sigma^x_2)), 
\end{equation}
where $D$ is the strength of spin-orbit coupling along the $z$ direction, hence $D=D_z$ and $(\sigma_x^i, \sigma_y^i, \sigma_z^i)$ are the Pauli operators for $ i$-th spin, $J_i$ are the spin interaction couplings. The Hamiltonian Eq. (\ref{Hxy1}) can be expressed in terms of  spin raising and lowering operators $\sigma^\pm$  as
\begin{equation}
H=( J+iD) \sigma_1^+ \sigma_2^-+(J-iD) \sigma_1^- \sigma_2^+ +\Delta( \sigma_1^+ \sigma_2^++ \sigma_1^- \sigma_2^-),
\end{equation}
where $J=(J_x+J_y)/2$ and $\Delta=(J_x-J_y)/2$ can be considered as the mean value  and anisotropy measures of spin interaction couplings respectively.  For simplicity, we have taken the value of $J$ to be 1. 
$\ell_\rho$ in Eq. (\ref{master}) is a non-unitary evolution of decoherence effect in the Lindblad form  in which $\gamma$ and $b_j$ are respectively the strength and the interaction operators of the system-environment coupling.  Eq. (\ref{master}) is a system of 16 linear differential equations which can be solved by standard methods. Differential equations are represented for dissipative, noisy and dephasing models in the Appendix. The interaction operators $b_j$  are defined as the spin raising and lowering operators $\sigma^\pm$ which can be modified for different models of decoherence and act on $j$-th spin. For instance $\sigma^+_1=\sigma^+\otimes I_{2\times2}$ and $\sigma^+_2= I_{2\times2}\otimes\sigma^+$, where $I_{2\times2}$ is an identity  $2\times2$ matrix \cite{label31}.

For a dissipative environment, the system is coupled with a thermal bath at zero temperature. It can be considered as an energy decay of a qubit due to spontaneous emission. The non-unitary part of Eq. (\ref{master}) reads 
\begin{equation}\label{el-diss}
\ell_\rho=\frac{\gamma}{2} \sum_{j=1,2}(2 \sigma_j^{-}\rho(t) \sigma_j^{+}- \sigma_j^{+} \sigma_j^{-} \rho(t)- \rho(t) \sigma_j^{+} \sigma_j^{-} ),
\end{equation}
and the nonzero elements of density matrix of $\rho(t)$ are given by 

\begin{subequations}\label{diss}
\begin{align}
\rho_{11}(t)&=\frac{e^{-2\gamma t}}{4|\omega|^2}((1-p)\gamma^2+2ie^{\omega t}(e^{4i\Delta t}-1)\gamma \Delta+4\Delta^2(e^{2\gamma t}-p)),\\
\rho_{14}(t)&=\rho^*_{41}(t)=\frac{e^{-\omega^* t}}{4|\omega|^2}\gamma((e^{4i\Delta t}-1)\gamma+2i\Delta(e^{4i\Delta t}-2e^{\omega^*t}+1)),\\
\rho_{22}(t)&=\frac{e^{-2\gamma t}}{4\nu^2|\omega|^2}[-De^{t(-2i\nu +\gamma)}(e^{4i\nu t}-1)(i\nu p|\omega|^2)+
\nonumber\\
&~\hspace*{2cm}\nu^2((p-1+e^{\omega t}+e^{\omega^* t})\gamma^2)+4\Delta^2(e^{2\gamma t}+p)],\\
\rho_{33}(t)&=\frac{e^{-2\gamma t}}{4\nu^2|\omega|^2}[De^{t(-2i\nu +\gamma)}(e^{4i\nu t}-1)(i\nu p|\omega|^2)+
\nonumber\\
&~\hspace*{2cm}\nu^2((p-1+e^{\omega t}+e^{\omega^* t})\gamma^2)
+4\Delta^2(e^{2\gamma t}+p)],\\
\rho_{23}(t)&=\rho^*_{32}(t)=\frac{e^{-\gamma t}p(iJ+D\cos(2\nu t))}{2(D+iJ)},\\
\rho_{44}(t)&=\frac{e^{-2\gamma t}}{4|\omega|^2}[\gamma^2(1+4e^{2\gamma t}-2e^{\omega t}-2e^{\omega^* t}-p)-2ie^{\omega t}(e^{4i\Delta t}-1)\gamma \Delta
\nonumber\\
&~\hspace*{2cm}+4\Delta^2(e^{2\gamma t}-p)],
\end{align}
\end{subequations}
 $\nu=\sqrt{D^2+J^2}$ and $\omega=\gamma-2i\Delta$.

For the infinite temperature or noisy environment, the non-unitary part of Eq. (\ref{master}) can be rewrite as follows:
\begin{align}\label{el-nois}
\ell_\rho&=\frac{\gamma}{2}\sum_{j=1,2}[(2 \sigma_j^{-}\rho(t) \sigma_j^{+}- \sigma_j^{+} \sigma_j^{-} \rho(t)- \rho(t) \sigma_j^{+} \sigma_j^{-} )
\nonumber \\
&+(2 \sigma_j^{+}\rho(t) \sigma_j^{-}- \sigma_j^{-} \sigma_j^{+} \rho(t)- \rho(t) \sigma_j^{-} \sigma_j^{+} )],
\end{align}
and the nonzero elements of density matrix of $\rho(t)$ read
\begin{subequations}\label{nois}
\begin{align}
\rho_{11}(t)&=\rho_{44}(t)=\frac{1-pe^{-4\gamma t}}{4},\\
\rho_{14}(t)&=\rho^*_{41}(t)=0,\\
\rho_{22}(t)&=\frac{1}{4}[1+p(e^{-4\gamma t}+\frac{D}{i\nu}(e^{2t(i\nu-\gamma)}-e^{-2t(i\nu+\gamma)}))],\\
\rho_{33}(t)&=\frac{1}{4}[1+p(e^{-4\gamma t}-\frac{D}{i\nu}(e^{2t(i\nu-\gamma)}-e^{-2t(i\nu+\gamma)}))],\\
\rho_{23}(t)&=\rho^*_{32}(t)=\frac{pe^{-2\gamma t}}{2(D+iJ)}(iJ+D\cos(2\nu t)).
\end{align}
\end{subequations}

Another model of decoherence which has no classical counterpart is dephasing. In this case, without losing energy, phase (quantum information) is damaged, for example by scattering. The non-unitary part of Eq. (\ref{master}) reads
\begin{equation}\label{el-deph}
\ell_\rho=\frac{\gamma}{2}\sum_{j=1,2} (2 \sigma_j^{+}\sigma_j^{-}\rho(t) \sigma_j^{+}\sigma_j^{-}- \sigma_j^{+} \sigma_j^{-} \rho(t)- \rho(t) \sigma_j^{+} \sigma_j^{-} ),
\end{equation}
and the nonzero elements of density matrix of $\rho(t)$ are given by 
\begin{subequations}\label{deph}
\begin{align}
\rho_{11}(t)&=\rho_{44}(t)=\frac{1-p}{4},\\
\rho_{14}(t)&=\rho^*_{41}(t)=0,\\
\rho_{22}(t)&=\frac{1}{4}[1+p(1+\frac{4De^{-t(\gamma+K)/2}(e^{Kt}-1))]}{K},\\
\rho_{33}(t)&=\frac{1}{4}[1+p(1-\frac{4De^{-t(\gamma+K)/2}(e^{Kt}-1))]}{K},\\
\rho_{23}(t)&=\rho^*_{32}(t)=\frac{e^{-\gamma t}p}{2(D+iJ)}(iJ+De^{\gamma t/2}(\cosh(Kt/2)-\frac{\gamma \sinh(Kt/2)}{K})),
\end{align}
\end{subequations}

where $K=\sqrt{\gamma^2-16(D^2+J^2)}$.

\section{ Quantum Correlation and Entanglement Measures} 

Concurrence ($C$) as a measure of entanglement for a two-qubit states is defined as \cite{label25}
\begin{equation}\label{con1}
 C(\rho)=max(0,\sqrt{\lambda_1}-\sqrt{\lambda_2}-\sqrt{\lambda_3}-\sqrt{\lambda_4}),
\end{equation}
where $\lambda_i$ are the eigenvalues, in decreasing order, of the matrix
\begin{equation}\label{zita}
 \xi=\rho^*(\sigma_y^A\otimes\sigma_y^B)\rho(\sigma_y^A\otimes\sigma_y^B),
\end{equation}
$\rho^*$ is the complex conjugate of $\rho$ and $\sigma_y$ is the Pauli matrix in the $y$ direction.
For the density matrix of  equations (\ref{diss}), (\ref{nois}) and (\ref{deph}) concurrence is given by
\begin{equation}\label{con}
 C(\rho)=2max [ 0,|\rho_{23}|-\sqrt{\rho_{11}\rho_{44}},|\rho_{14}|-\sqrt{\rho_{22}\rho_{33}}].
\end{equation}

Quantum discord $(QD)$ as a measure of quantum correlation for two-qubit systems is defined as the difference between the quantum mutual information, $I( \rho)$, and the classical correlation($CC$) \cite{label26,label27,label28,label29}
\begin{equation}\label{qd1}
 QD( \rho)=I( \rho)-CC(\rho),
\end{equation}
  the quantum mutual information is defined as
\begin{equation}\label{iro}
 I( \rho)=S(\rho^A)+S(\rho^B)-S(\rho) 
\end{equation}
 where $S(\rho)=-Tr(\rho \log_2\rho)$, and $\rho^A(\rho^B)$ is the reduced state of subsystem $A(B)$ and classical correlation can be defined as follows:
\begin{equation}\label{clc1}
 CC( \rho)=S( \rho^{A})-min(S(\rho|\{B_i\})),
\end{equation}
where $S(\rho|\{B_i\})$ is the conditional entropy and minimization is taken over all possible Von Neumann measurements on subsystem {\it B}. The quantum conditional entropy is defined as
\begin{equation}\label{condition}
S(\rho|\{B_i\})=\sum_{i=0}^1 p_i S(\rho_i),
\end{equation}
where
\begin{equation}
 \rho_i=\frac{1}{p_i}(I_A\otimes B_i)\rho(I_A\otimes B_i)
\end{equation}
is the density operator after obtaining the outcome {\it i} of  subsystem {\it B}, with the probability
\begin{equation}
 p_i=Tr \left[ (I_A\otimes B_i)\rho(I_A\otimes B_i) \right].
\end{equation}

We use the method given by M. Ali et al. \cite{label29} to minimize $S(\rho|\{B_i\})$. As a result, $QD(\rho)$ and $CC(\rho)$  for the density operators, Eqs. (\ref{diss}), (\ref{nois}) and (\ref{deph}) are respectively given by
\begin{equation}\label{qd2}
QD(\rho)=S(\rho^B)-S(\rho)+min[S_1, S_2, S_3,S_4,S_5],
\end{equation}
\begin{equation}\label{clc2}
CC(\rho)=S(\rho^A)-min[S_1, S_2, S_3,S_4,S_5],
\end{equation}
 where $S_1=p_0h(\theta_0)+p_1h(\theta_1)$, $S_2=h(\theta_2)$, $S_3=h(\theta_3)$, $S_4=h(\theta_4)$ and $S_5=h(\theta_5)$, where $h(\theta)=-\frac{1-\theta}{2}\log_2{\frac{1-\theta}{2}}-\frac{1+\theta}{2}\log_2{\frac{1+\theta}{2}}$ and according to the matrix elements, $\theta_i$ are expressed as 
\begin{equation}
\theta_0=\left|\frac{\rho_{11}-\rho_{33}}{\rho_{11}+\rho_{33}}\right|,
\end{equation}
\begin{equation}
\theta_1=\left|\frac{\rho_{22}-\rho_{44}}{\rho_{22}+\rho_{44}}\right|,
\end{equation}
\begin{equation}
 \theta_{2,3}=2\sqrt{|\rho_{14}|^2+|\rho_{23}|^2 \pm 2Re(\rho_{14}\rho^*_{23})+\frac{1}{4}(\rho_{11}+\rho_{22}-\rho_{33}-\rho_{44})^2},
\end{equation}
\begin{equation}
 \theta_{4,5}=2\sqrt{|\rho_{14}|^2+|\rho_{23}|^2 \pm 2Im(\rho_{14}\rho^*_{23})+\frac{1}{4}(\rho_{11}+\rho_{22}-\rho_{33}-\rho_{44})^2},
\end{equation}
and their corresponding probabilities are $p_0=\rho_{11}+\rho_{33}$ and $p_1=\rho_{22}+\rho_{44}$. 

\section{ Comparison} 
\subsection{Without decoherence}
At first, we assume that $\gamma=0$ for which there is no decoherence effect. In this case, when there is no spin-orbit interaction, $D=0$, quantum and classical correlations, as well as entanglement, have  constant values depending on the initial state. For instance if $p=1/2$, then $QD=0.26$, $CC=0.19$ and $C=0.25$. They begin to oscillate with nonzero values of $D$. We have plotted time evolution of concurrence, quantum discord and classical correlation in Fig. 1 for $p=0.5$ and different values of $D$. From Fig. 1, we see that there is a critical value of $D$ for which concurrence vanishes periodically  and ESD happens while quantum discord and classical correlation remain nonzero with the same oscillations. It is obvious that with the growth of spin-orbit interaction the frequency of quantifiers increases. In general the time frequencies of quantifiers are the same for all correlations, whereas the amplitude of classical correlation is less than the amplitude of quantum discord and  concurrence. Note that from Eq. (\ref{diss}) the quantifiers are independent of $\Delta$ if $\gamma=0$.

\subsection{With decoherence}
In this section we compare the dynamics of  entanglement, quantum and classical correlations in the presence of decoherence. In this case, we are interested in the behavior of correlations at infinite time where these quantities obtain constant values. We assume that the strength of the system-environment coupling is weak and $\gamma=1/2$ for all cases. At first we investigate the case of  dissipation. When  $t\rightarrow\infty$  the nonzero elements of density operator Eq. (\ref{diss}) are reduced as 
\begin{subequations}\label{infi}
\begin{align}
\rho_{11}&=\rho_{22}=\rho_{33}=\Delta^2/(4\Delta^2+\gamma^2),\\ 
\rho_{14}&=\rho^*_{41}=-i\gamma \Delta/(4\Delta^2+\gamma^2),\\
\rho_{44}&=(\Delta^2+\gamma^2)/(4\Delta^2+\gamma^2).
\end{align}
\end{subequations}

It is obvious that the density matrix is independent of $D$ and $p$ at infinite time. So for simplicity we assume $D=0$ and $p=0.5$. From equations (\ref{qd2}), (\ref{clc2}) and (\ref{infi}) quantum discord and classical correlation exist for nonzero values of $\Delta$. But concurrence is reduced to
\begin{equation}\label{cinfi}
C(t\rightarrow\infty)=2max[0,\frac{(\gamma \Delta-\Delta^2)}{(4\Delta^2+\gamma^2)}],
\end{equation}
therefore entanglement vanishes for $\Delta\geq\gamma$ and $\Delta=0$. These results indicate that correlations can be controlled by the anisotropic parameter, $\Delta$, and quantum and classical correlations are more stable than entanglement.

  We have plotted correlation measures as a function of time and $\Delta$ in Fig. 2. From Figs. 2(a) and (b), quantum and classical correlations have nonzero values for $\Delta>0$, but in Fig. 2(c) entanglement vanishes for  $\Delta\geq1/2$ and $\Delta=0$  as we expected.  We also have plotted the dynamics of correlation measures  for different values of $\Delta$ in Fig. 3. If $\Delta=0$, all correlations go to zero quickly as depicted in Fig. 3(a). For nonzero and weak anisotropic  values of $\Delta$ the correlations obtain constant values at infinite time as shown in figures 3(b) and 3(c). For example from Eq. (\ref{cinfi}) if $\Delta=0.2$,  the value of concurrence is $0.29$ after occurrence entanglement sudden death. By substituting (\ref{infi})  in (\ref{qd2}) and (\ref{clc2}) we obtain $QD=CC=0.21$ for $\Delta=0.2$ without vanishing as shown in Fig. 3(b).

 In the case of  strong value of anisotropic parameter, $\Delta=0.8$,  concurrence vanishes abruptly,  while quantum discord and classical correlation obtain the same stable value $QD=CC=0.06$ as depicted in Fig. 3(d). 

We are also interested in the behavior of correlations when the state is completely mixed; that is at $p=0$. Since all correlations depend on initial condition $p$, therefore there are not any correlations at $t=0$; but, they revive after awhile for weak values of $\Delta$ as shown in Figs. 4(a) and 4(b). Contrary to the other correlations, quantum entanglement does not revive for $\Delta=0.8$ and remains zero for ever as depicted in Fig. 4(c). Our numerical results show that the dynamics of $QD$ and $CC$ follow the same dynamics for $p=0$ (Fig. 4). Since the results are based on both analytical and numerical procedures (e.g. Eq. (\ref{qd2})), it is difficult to show how $QD$ and $CC$ behave in such a way. However, for instance we consider the density matrix, Eq. (\ref{diss}) at $t=2$ for $p=0$ and $\Delta=0.4$; therefore the nonzero elements of the density matrix in Eq. (\ref{diss}) are given by
\begin{subequations}\label{inst}
\begin{align}
\rho_{11}&=0.11,\\
 \rho_{22}&=\rho_{33}=0.17,\\
\rho_{14}&=\rho^*_{41}=-0.17 i,\\
\rho_{44}&=0.55.
\end{align}
\end{subequations}
According to Eq. (\ref{inst}) we have $S(\rho^A)=S(\rho^B)=0.85$, $S_1=0.83$, $S_2=S_3=S_4=S_5=0.75$, $S(\rho)=1.5$ and hence $S(\rho)=2min[S_1, S_2, S_3,S_4,S_5]$. By substituting these values in (\ref{qd2}) and (\ref{clc2}) we obtain $QD=CC=0.1$.  Quantum discord may be greater, less and equal to classical correlation as depicted in Fig. 3. They also happen for mixed and pure states as shown in references \cite{label28,label29,label30}.

In the noisy and dephasing environment, their corresponding density operators are independent of the parameter $\Delta$, from equations (\ref{nois}) and (\ref{deph}) respectively. When $t\rightarrow\infty$, the noisy environment induces the system to a mixed state and the density operator, Eq. (\ref{nois}) is reduced to $\rho_{11}=\rho_{22}=\rho_{33}=\rho_{44}=1/4$; therefore from (\ref{con}),  (\ref{qd2}) and  (\ref{clc2}) all correlations go to zero quickly,  however concurrence decays more quickly than the other quantifiers as depicted in Fig. 5(a).

 In the case of dephasing environment  Eq. (\ref{deph}) is reduced to $\rho_{11}=\rho_{44}=(1-p)/4$ and $\rho_{22}=\rho_{33}=(1+p)/4$ after awhile. In spite of dissipative and noisy environment, in this case the density operator depends on initial state, $p$. Since dephasing environment just destroys quantum information, classical correlation exists for nonzero values of $p$. For  instance from Eq. (\ref{clc2}) $CC=0.19$ for $p=1/2$ and there is neither entanglement  nor quantum discord from Eqs. (\ref{con}) and (\ref{qd2}) respectively as displayed in Fig. 5(b). It is also clear that entanglement decays more quickly than quantum discord as it happened in the noisy case.

\section{Conclusions} 
In this work we consider a two-qubit state in Werner state which evolves in Heisenberg XY model with the DM interaction under various system-environment couplings. We evaluate  and compare entanglement, quantum and classical correlations. We show that in the absence of decoherence effects, there is a critical value of DM interaction for which quantum entanglement may vanish, while quantum and classical correlations have nonzero values. In the presence of a dissipative environment,  the anisotropic parameter can control the behavior of correlations and prevent them from decaying, while in the isotropic case the correlations go to zero quickly. Our numerical results show that quantum and classical correlations reach to the same value at infinite time. They also evolve in the same way when the initial state is completely mixed.

 The noisy environment enforces the correlations to decay abruptly. Classical correlation just depends on initial state and does not decay in the dephasing environment, but quantum information is destroyed by this kind of environment,  thus quantum discord, as well as entanglement decays to zero. 

As a result, it is shown that the quantum and classical correlations are more robust than the quantum entanglement in the absence and presence of decoherence.

\section*{Acknowledgments}
We would like to thank Haitham Zaraket and Mojtaba Jafarpour for reading this article and making useful comments. Also, M. R. Pourkarimi would like to thank Mazhar Ali for private communication.


\subsection*{Appendix: Differential Equations of Eq.(\ref{master}) For Dissipative, Noisy and Dephasing Models}
\renewcommand{\theequation}{A--\arabic{equation}}
\setcounter{equation}{0}
1. Differential equations for the dissipative environment:

\begin{align}
\rho _{11}'[t]&=-2 \gamma  \rho _{11}[t]+i \Delta  \left(\rho _{14}[t]-\rho _{41}[t]\right),\\
\rho _{12}'[t]&=-\frac{3}{2} \gamma  \rho _{12}[t]+(D+i J) \rho _{13}[t]-i \Delta  \rho _{42}[t],\\
\rho _{13}'[t]&=-(D-i J) \rho _{12}[t]-\frac{3}{2} \gamma  \rho _{13}[t]-i \Delta  \rho _{43}[t],\\
\rho _{14}'[t]&=-\gamma  \rho _{14}[t]+i \Delta  \left(\rho _{11}[t]-\rho _{44}[t]\right),\\
\rho _{21}'[t]&=-\frac{3}{2} \gamma  \rho _{21}[t]+i \Delta  \rho _{24}[t]+(D-i J) \rho _{31}[t],\\
\rho _{22}'[t]&=\gamma  \rho _{11}[t]-\gamma  \rho _{22}[t]+(D+i J) \rho _{23}[t]+(D-i J) \rho _{32}[t],\\
\rho _{23}'[t]&=-\gamma  \rho _{23}[t]-(D-i J) \left(\rho _{22}[t]-\rho _{33}[t]\right),\\
\rho _{24}'[t]&=\gamma  \rho _{13}[t]+i \Delta  \rho _{21}[t]-\frac{1}{2} \gamma  \rho _{24}[t]+(D-i J) \rho _{34}[t],\\
\rho _{31}'[t]&=-(D+i J) \rho _{21}[t]-\frac{3}{2} \gamma  \rho _{31}[t]+i \Delta  \rho _{34}[t],\\
\rho _{32}'[t]&=-\gamma  \rho _{32}[t]-(D+i J) \left(\rho _{22}[t]-\rho _{33}[t]\right),\\
\rho _{33}'[t]&=\gamma  \rho _{11}[t]-(D+i J) \rho _{23}[t]-(D-i J) \rho _{32}[t]-\gamma  \rho _{33}[t],\\
\rho _{34}'[t]&=\gamma  \rho _{12}[t]-(D+i J) \rho _{24}[t]+i \Delta  \rho _{31}[t]-\frac{1}{2} \gamma  \rho _{34}[t],\\
\rho _{41}'[t]&=-\gamma  \rho _{41}[t]-i \Delta  \left(\rho _{11}[t]-\rho _{44}[t]\right),\\
\rho _{42}'[t]&=-i \Delta  \rho _{12}[t]+\gamma  \rho _{31}[t]-\frac{1}{2} \gamma  \rho _{42}[t]+(D+i J) \rho _{43}[t],\\
\rho _{43}'[t]&=-i \Delta  \rho _{13}[t]+\gamma  \rho _{21}[t]-(D-i J) \rho _{42}[t]-\frac{1}{2} \gamma  \rho _{43}[t],\\
\rho _{44}'[t]&=\gamma  \left(\rho _{22}[t]+\rho _{33}[t]\right)-i \Delta  \left(\rho _{14}[t]-\rho _{41}[t]\right).
\end{align}
2. Differential equations for the noisy environment:
\renewcommand{\theequation}{A--\arabic{equation}}
\setcounter{equation}{16}
\begin{align}
\rho _{11}'[t]&=\gamma  \left(-2 \rho _{11}[t]+\rho _{22}[t]+\rho _{33}[t]\right)+i \Delta  \left(\rho _{14}[t]-\rho _{41}[t]\right),\\ 
\rho _{12}'[t]&=-2 \gamma  \rho _{12}[t]+(D+i J) \rho _{13}[t]+\gamma  \rho _{34}[t]-i \Delta  \rho _{42}[t],\\
\rho _{13}'[t]&=-(D-i J) \rho _{12}[t]-2 \gamma  \rho _{13}[t]+\gamma  \rho _{24}[t]-i \Delta  \rho _{43}[t],\\
\rho _{14}'[t]&=-2 \gamma  \rho _{14}[t]+i \Delta  \left(\rho _{11}[t]-\rho _{44}[t]\right),\\
\rho _{21}'[t]&=-2 \gamma  \rho _{21}[t]+i \Delta  \rho _{24}[t]+(D-i J) \rho _{31}[t]+\gamma  \rho _{43}[t],\\
\rho _{22}'[t]&=(D+i J) \rho _{23}[t]+(D-i J) \rho _{32}[t]+\gamma  \left(\rho _{11}[t]-2 \rho _{22}[t]+\rho _{44}[t]\right),\\
\rho _{23}'[t]&=-2 \gamma  \rho _{23}[t]-(D-i J) \left(\rho _{22}[t]-\rho _{33}[t]\right),\\
\rho _{24}'[t]&=\gamma  \rho _{13}[t]+i \Delta  \rho _{21}[t]-2 \gamma  \rho _{24}[t]+(D-i J) \rho _{34}[t],\\
\rho _{31}'[t]&=-(D+i J) \rho _{21}[t]-2 \gamma  \rho _{31}[t]+i \Delta  \rho _{34}[t]+\gamma  \rho _{42}[t],\\
\rho _{32}'[t]&=-2 \gamma  \rho _{32}[t]-(D+i J) \left(\rho _{22}[t]-\rho _{33}[t]\right),\\
\rho _{33}'[t]&=-(D+i J) \rho _{23}[t]-(D-i J) \rho _{32}[t]+\gamma  \left(\rho _{11}[t]-2 \rho _{33}[t]+\rho _{44}[t]\right),\\
\rho _{34}'[t]&=\gamma  \rho _{12}[t]-(D+i J) \rho _{24}[t]+i \Delta  \rho _{31}[t]-2 \gamma  \rho _{34}[t],\\
\rho _{41}'[t]&=-2 \gamma  \rho _{41}[t]-i \Delta  \left(\rho _{11}[t]-\rho _{44}[t]\right),\\
\rho _{42}'[t]&=-i \Delta  \rho _{12}[t]+\gamma  \rho _{31}[t]-2 \gamma  \rho _{42}[t]+(D+i J) \rho _{43}[t],\\
\rho _{43}'[t]&=-i \Delta  \rho _{13}[t]+\gamma  \rho _{21}[t]-(D-i J) \rho _{42}[t]-2 \gamma  \rho _{43}[t],\\
\rho _{44}'[t]&=-i \Delta  \left(\rho _{14}[t]-\rho _{41}[t]\right)+\gamma  \left(\rho _{22}[t]+\rho _{33}[t]-2 \rho _{44}[t]\right).
\end{align}

3. Differential equations for the dephasing environment:
\renewcommand{\theequation}{A--\arabic{equation}}
\setcounter{equation}{32}
\begin{align}
\rho _{11}'[t]&=i \Delta  \left(\rho _{14}[t]-\rho _{41}[t]\right),\\
\rho _{12}'[t]&=-\frac{1}{2} \gamma  \rho _{12}[t]+(D+i J) \rho _{13}[t]-i \Delta  \rho _{42}[t],\\
\rho _{13}'[t]&=-(D-i J) \rho _{12}[t]-\frac{1}{2} \gamma  \rho _{13}[t]-i \Delta  \rho _{43}[t],\\
\rho _{14}'[t]&=-\gamma  \rho _{14}[t]+i \Delta  \left(\rho _{11}[t]-\rho _{44}[t]\right),\\
\rho _{21}'[t]&=-\frac{1}{2} \gamma  \rho _{21}[t]+i \Delta  \rho _{24}[t]+(D-i J) \rho _{31}[t],\\
\rho _{22}'[t]&=(D+i J) \rho _{23}[t]+(D-i J) \rho _{32}[t],\\
\rho _{23}'[t]&=-\gamma  \rho _{23}[t]-(D-i J) \left(\rho _{22}[t]-\rho _{33}[t]\right),\\
\rho _{24}'[t]&=i \Delta  \rho _{21}[t]-\frac{1}{2} \gamma  \rho _{24}[t]+(D-i J) \rho _{34}[t],\\
\rho _{31}'[t]&=-(D+i J) \rho _{21}[t]-\frac{1}{2} \gamma  \rho _{31}[t]+i \Delta  \rho _{34}[t],\\
\rho _{32}'[t]&=-\gamma  \rho _{32}[t]-(D+i J) \left(\rho _{22}[t]-\rho _{33}[t]\right),\\
\rho _{33}'[t]&=-(D+i J) \rho _{23}[t]-(D-i J) \rho _{32}[t],\\
\rho _{34}'[t]&=-(D+i J) \rho _{24}[t]+i \Delta  \rho _{31}[t]-\frac{1}{2} \gamma  \rho _{34}[t],\\
\rho _{41}'[t]&=-\gamma  \rho _{41}[t]-i \Delta  \left(\rho _{11}[t]-\rho _{44}[t]\right),\\
\rho _{42}'[t]&=-i \Delta  \rho _{12}[t]-\frac{1}{2} \gamma  \rho _{42}[t]+(D+i J) \rho _{43}[t],\\
\rho _{43}'[t]&=-i \Delta  \rho _{13}[t]-(D-i J) \rho _{42}[t]-\frac{1}{2} \gamma  \rho _{43}[t],\\
\rho _{44}'[t]&=-i \Delta  \left(\rho _{14}[t]-\rho _{41}[t]\right),
\end{align}
where prime ($\prime$) is employed for the time derivative. 

\begin{figure}[h]
\includegraphics[width=0.5\textwidth]{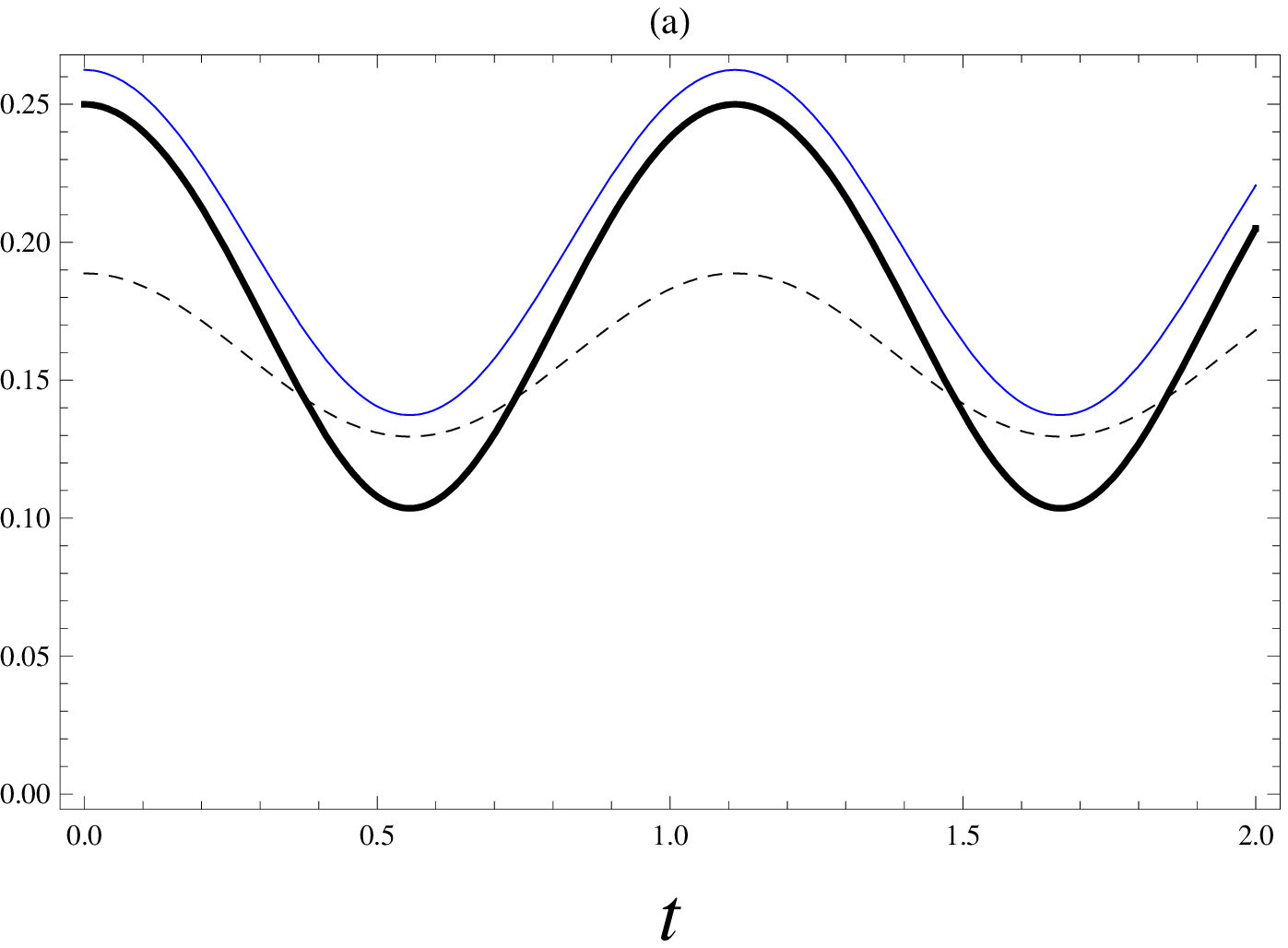}
\includegraphics[width=0.5\textwidth]{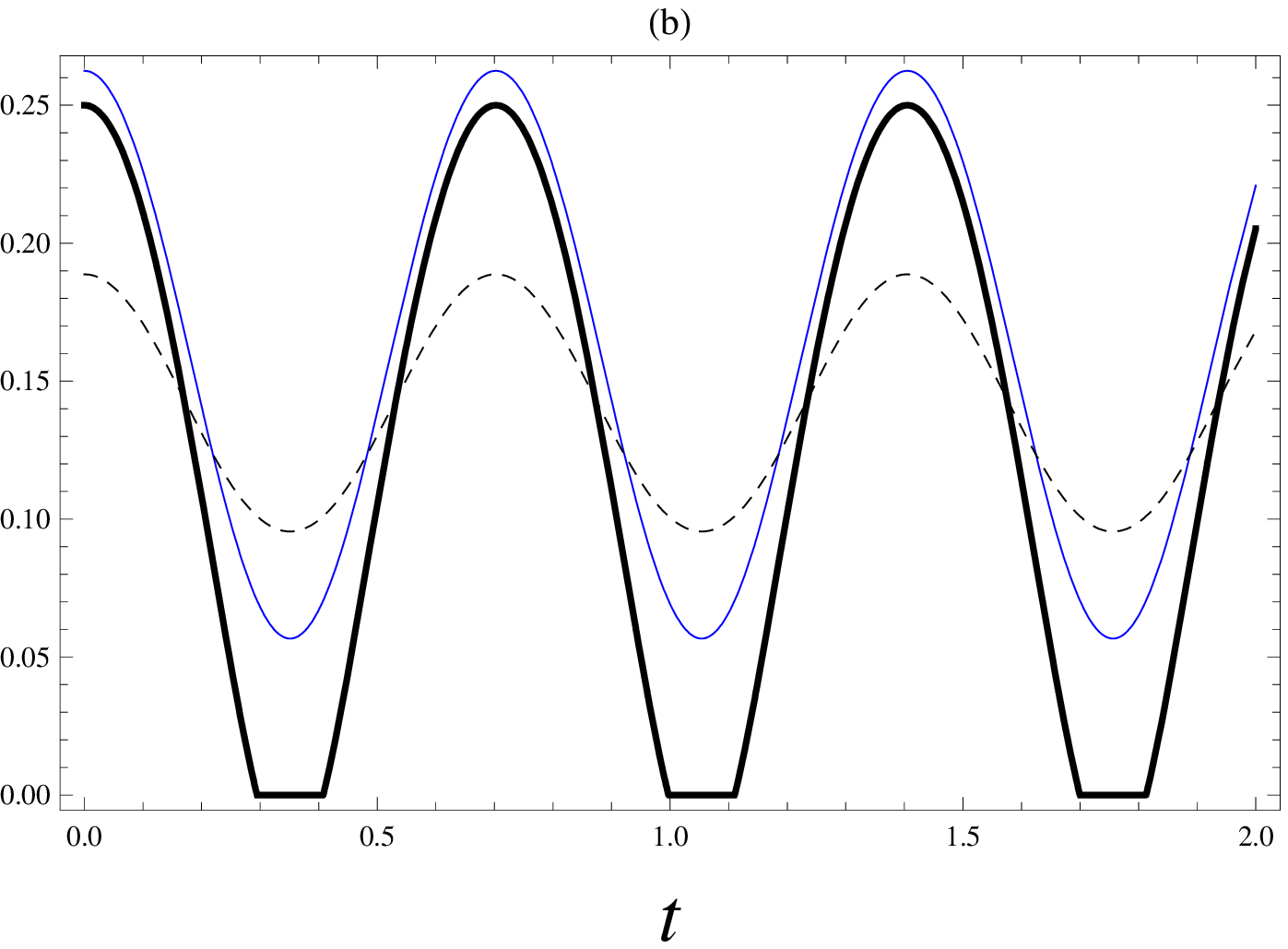}
\includegraphics[width=0.5\textwidth]{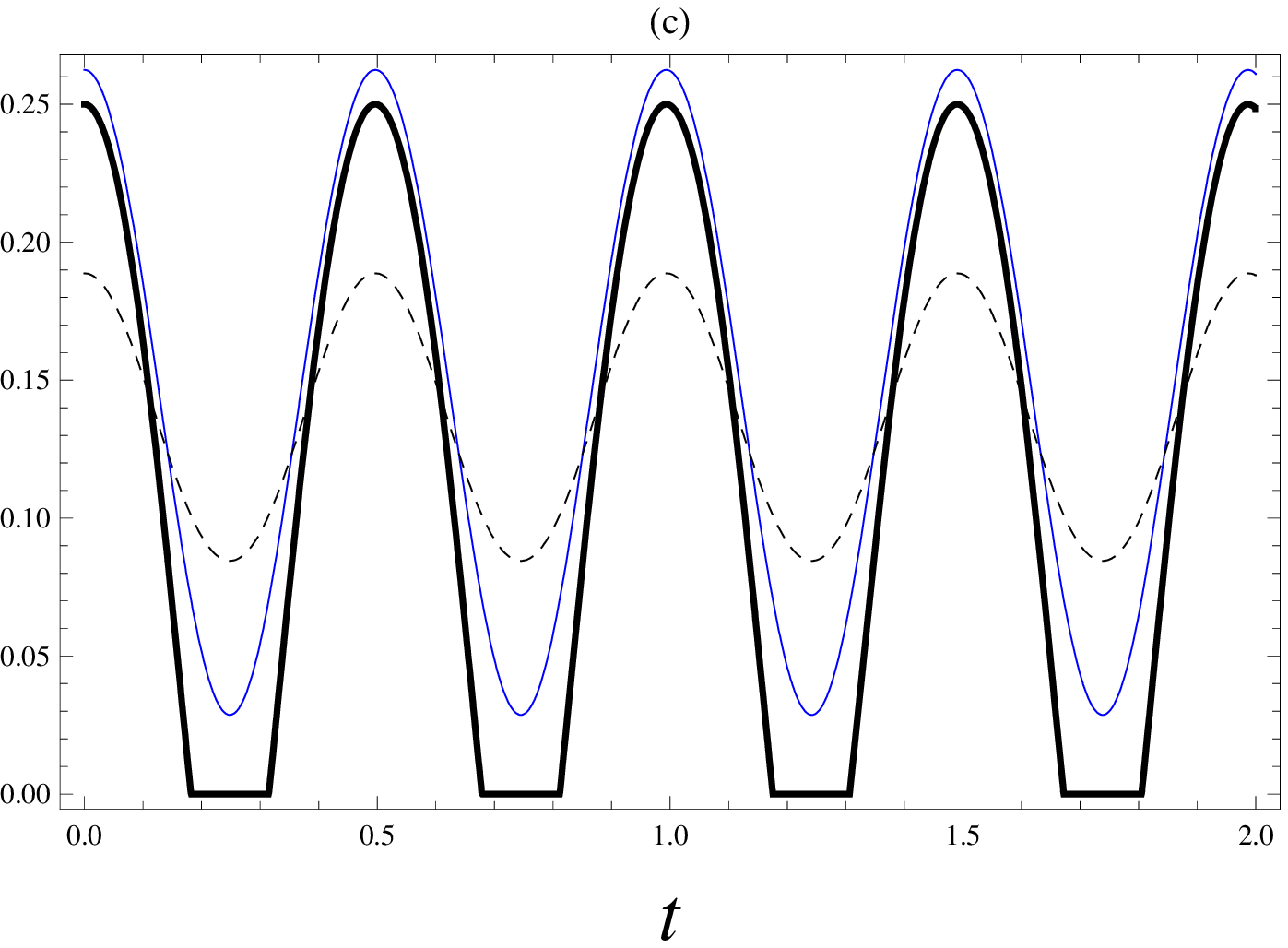}
\includegraphics[width=0.5\textwidth]{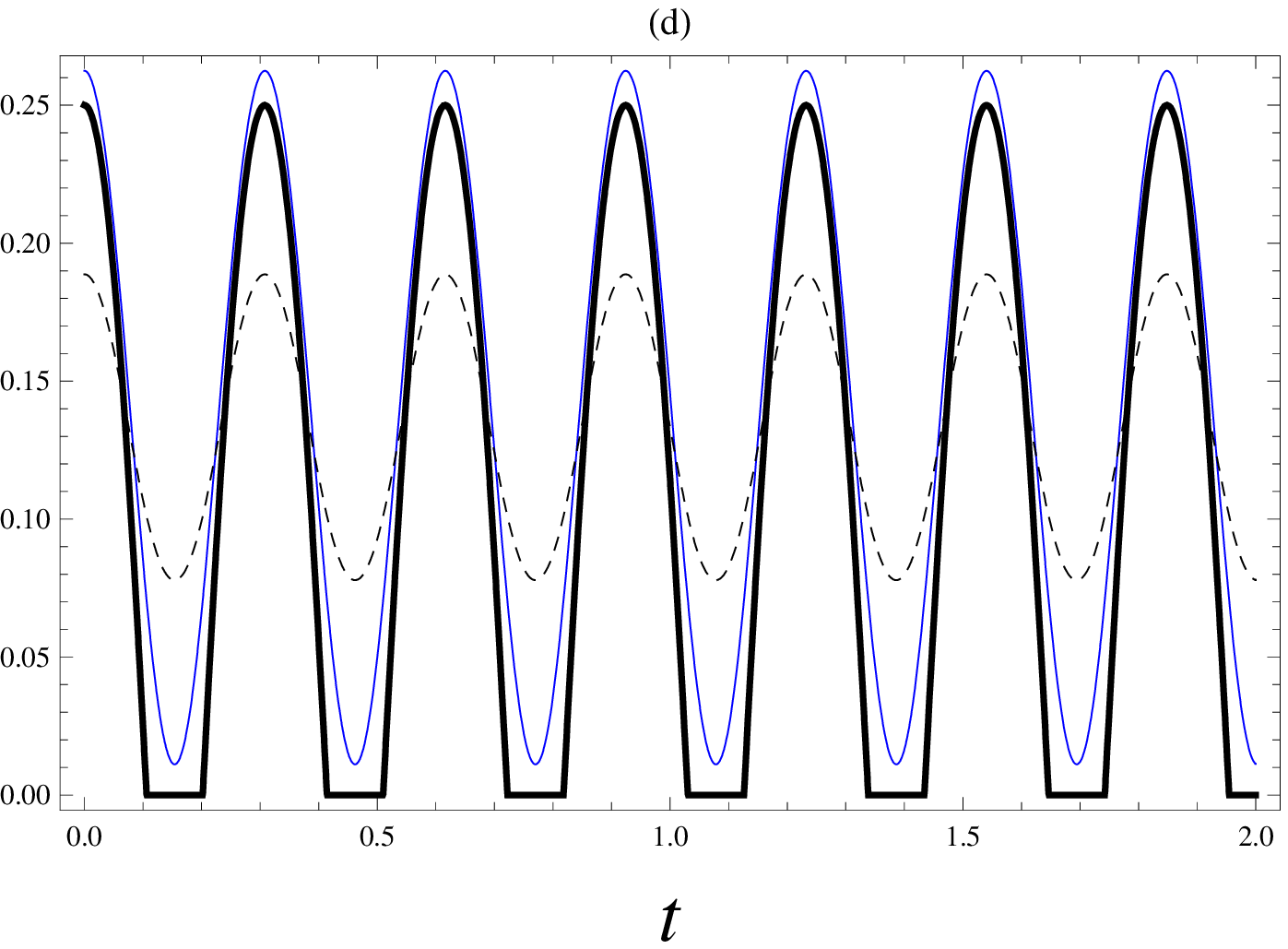}
\caption{Concurrence (thick solid line), quantum discord (thin solid line) and classical correlation (dashed line) as a function of time, in the absence of decoherence ($\gamma=0$) for $p=0.5$, (a) $D=1$, (b) $D=2$, (c) $D=3$ and  (d) $D=5$.} \label{figure1}
\end{figure}

\begin{figure}[ht]
\includegraphics[width=0.5\textwidth]{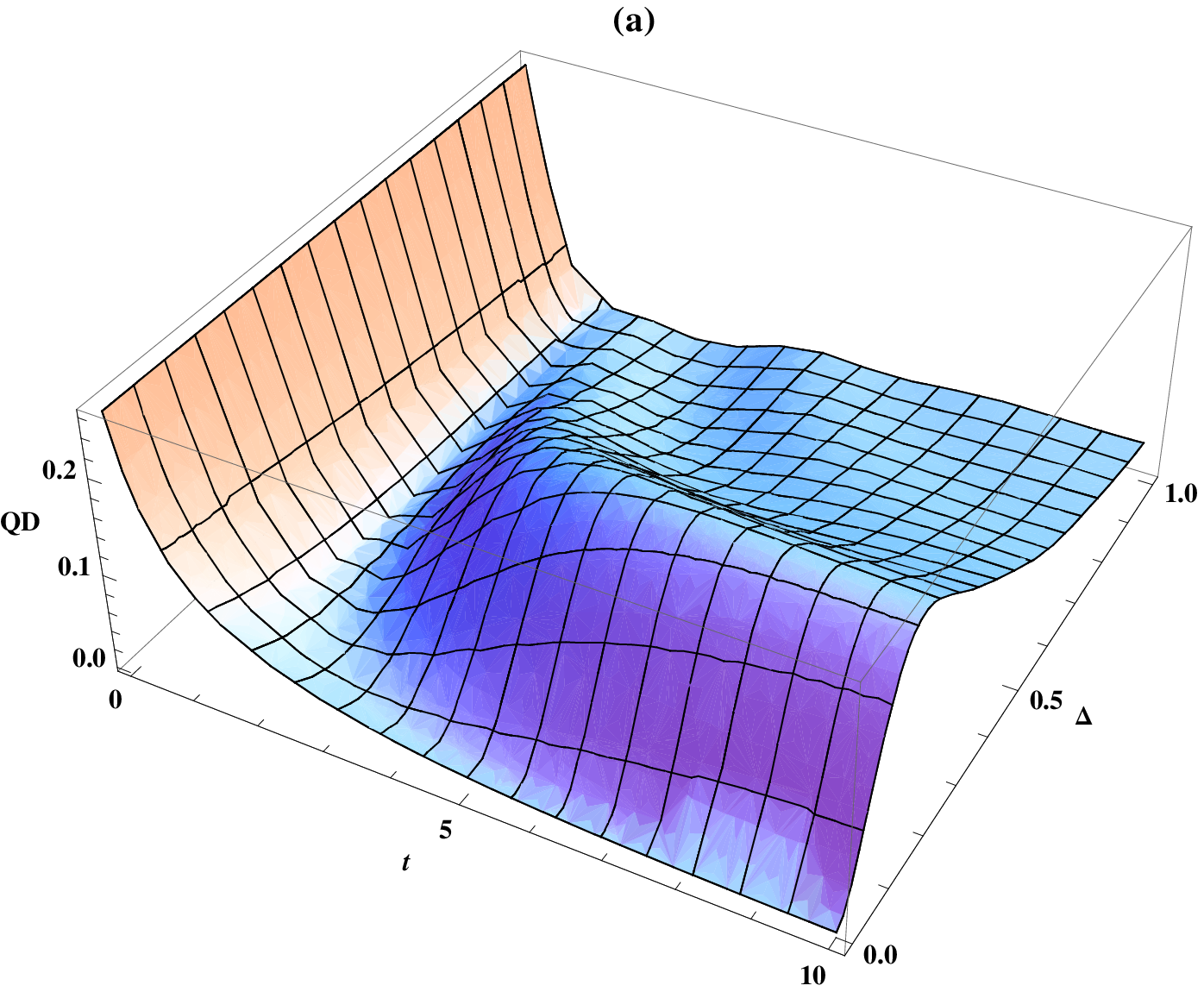}
\includegraphics[width=0.5\textwidth]{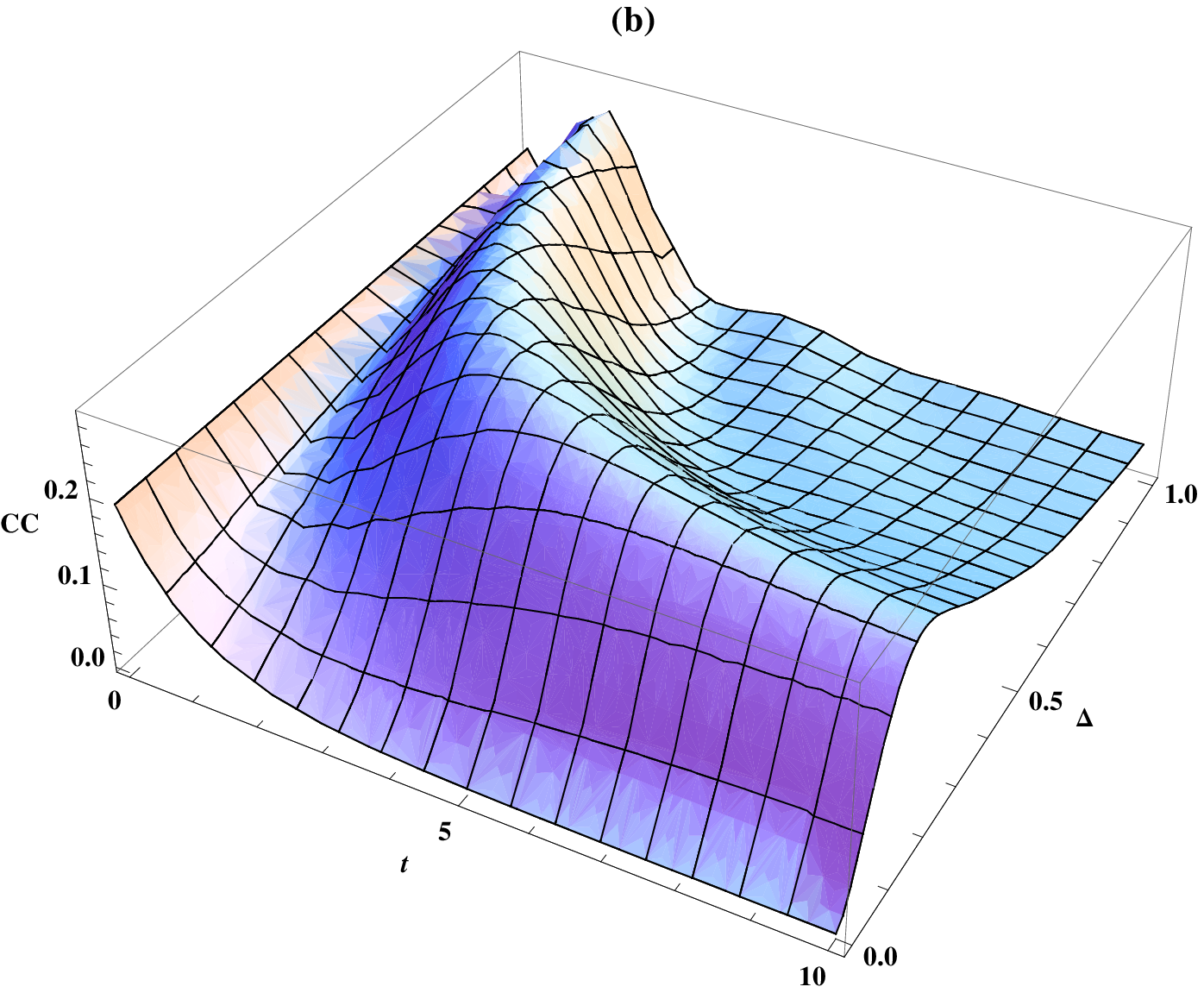}
\includegraphics[width=0.5\textwidth]{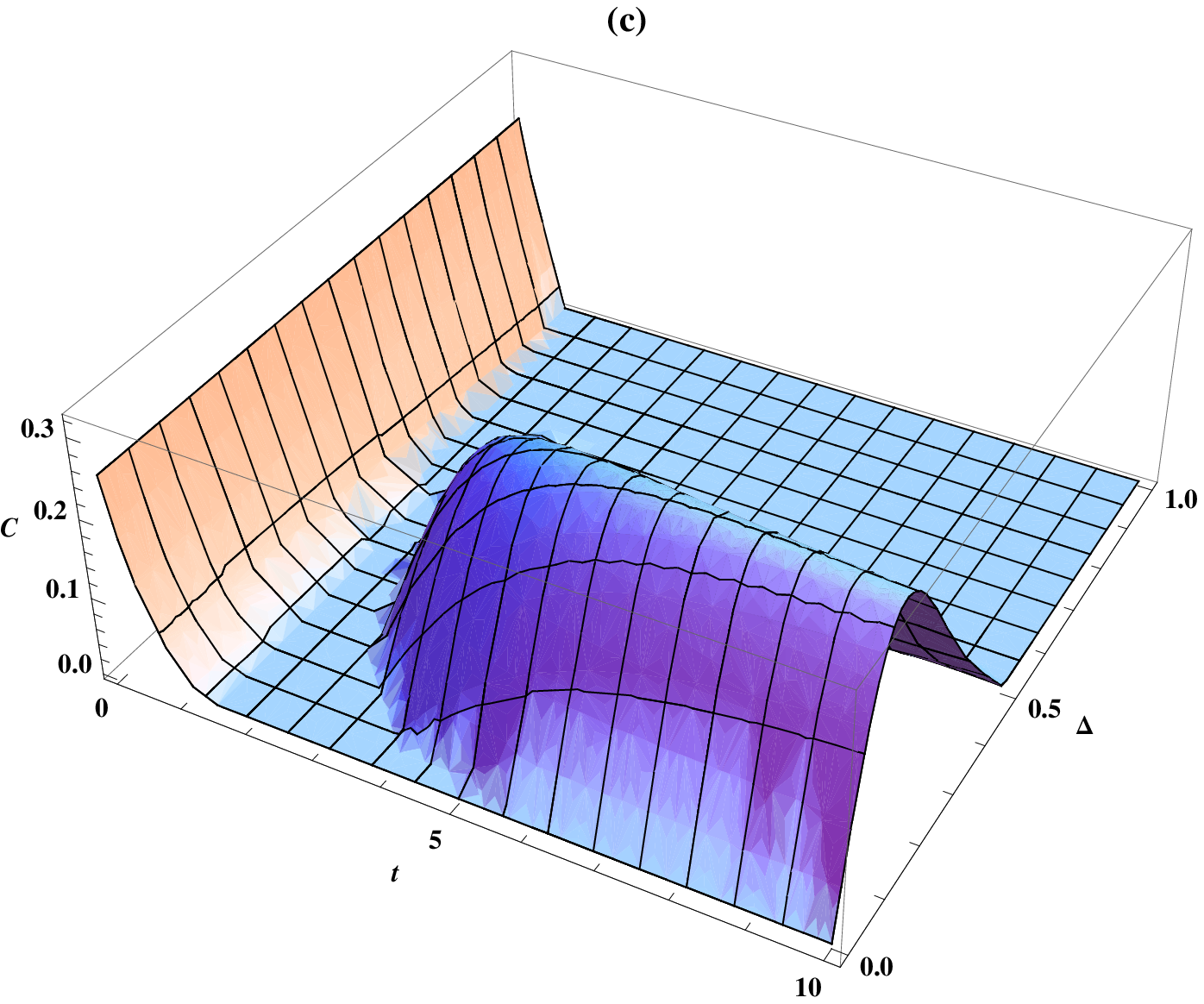}
\caption{(a) Quantum discord ($QD$), (b) classical correlation ($CC$) and (c) concurrence ($C$) versus time and anisotropic parameter ($\Delta$) for $p=0.5$, $\gamma=0.5$ and $D=0$ in the case of dissipative environment.} \label{figure2}
\end{figure}

 \begin{figure}
\includegraphics[width=0.5\textwidth]{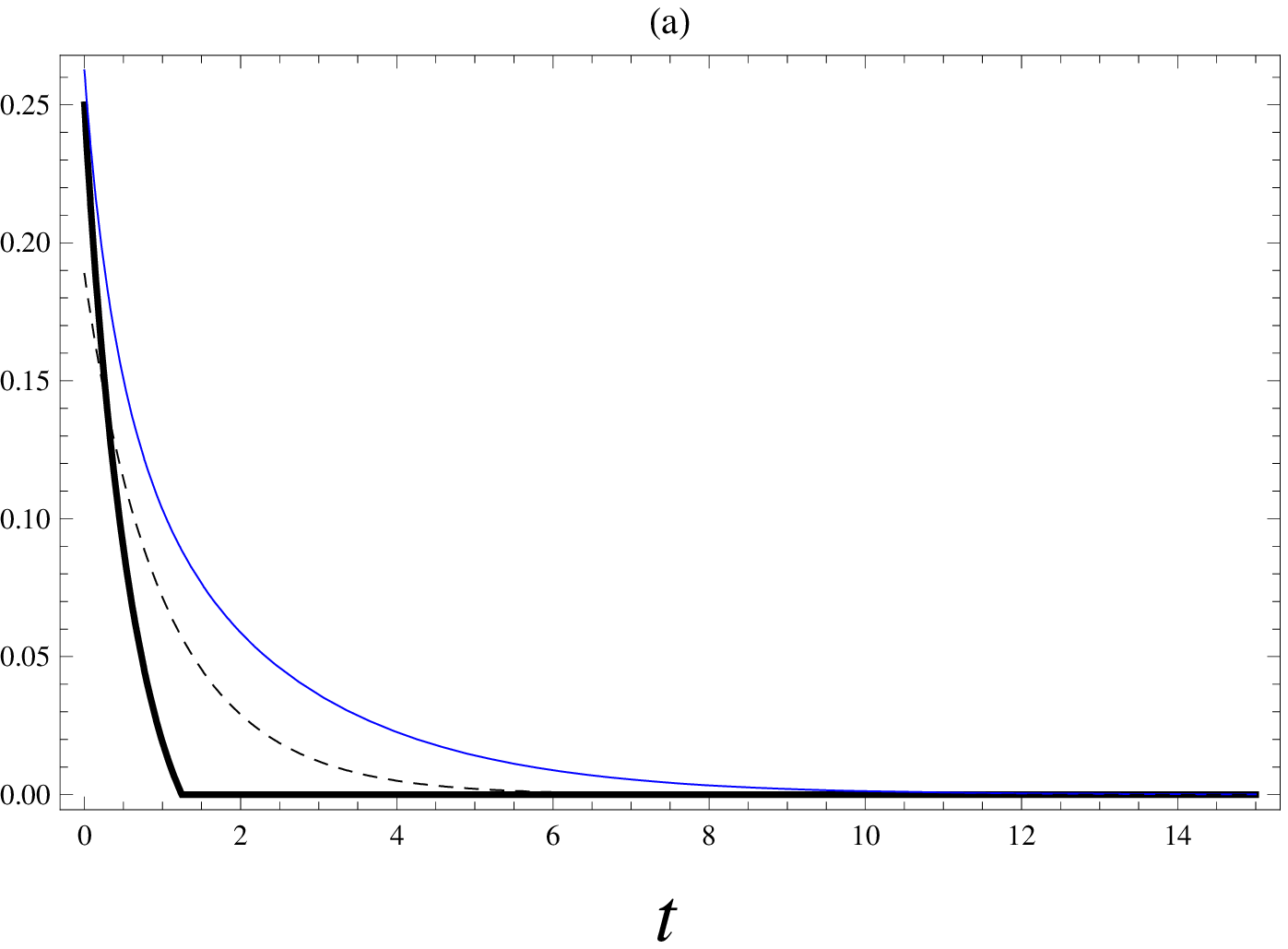}
\includegraphics[width=0.5\textwidth]{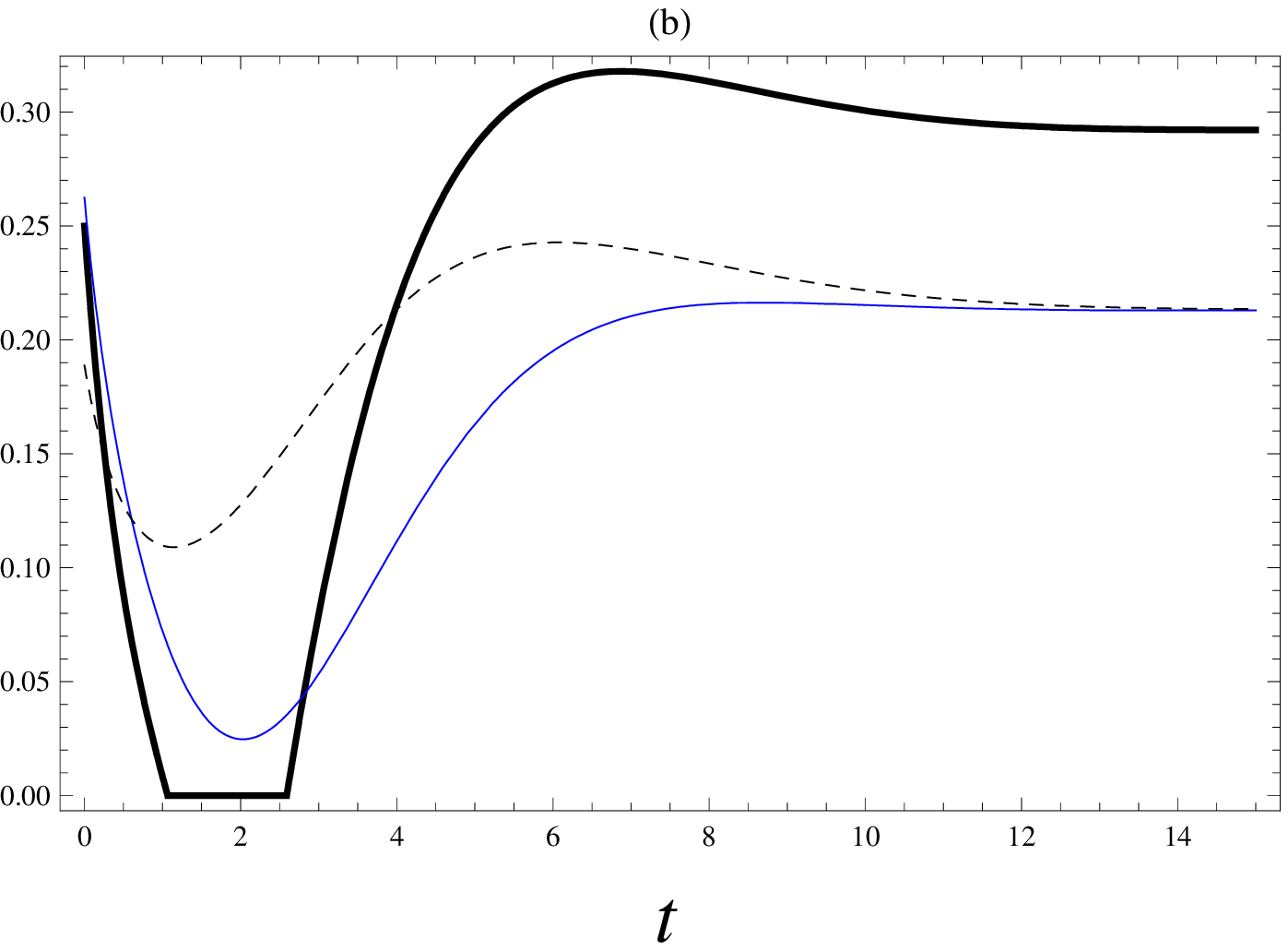}
\includegraphics[width=0.5\textwidth]{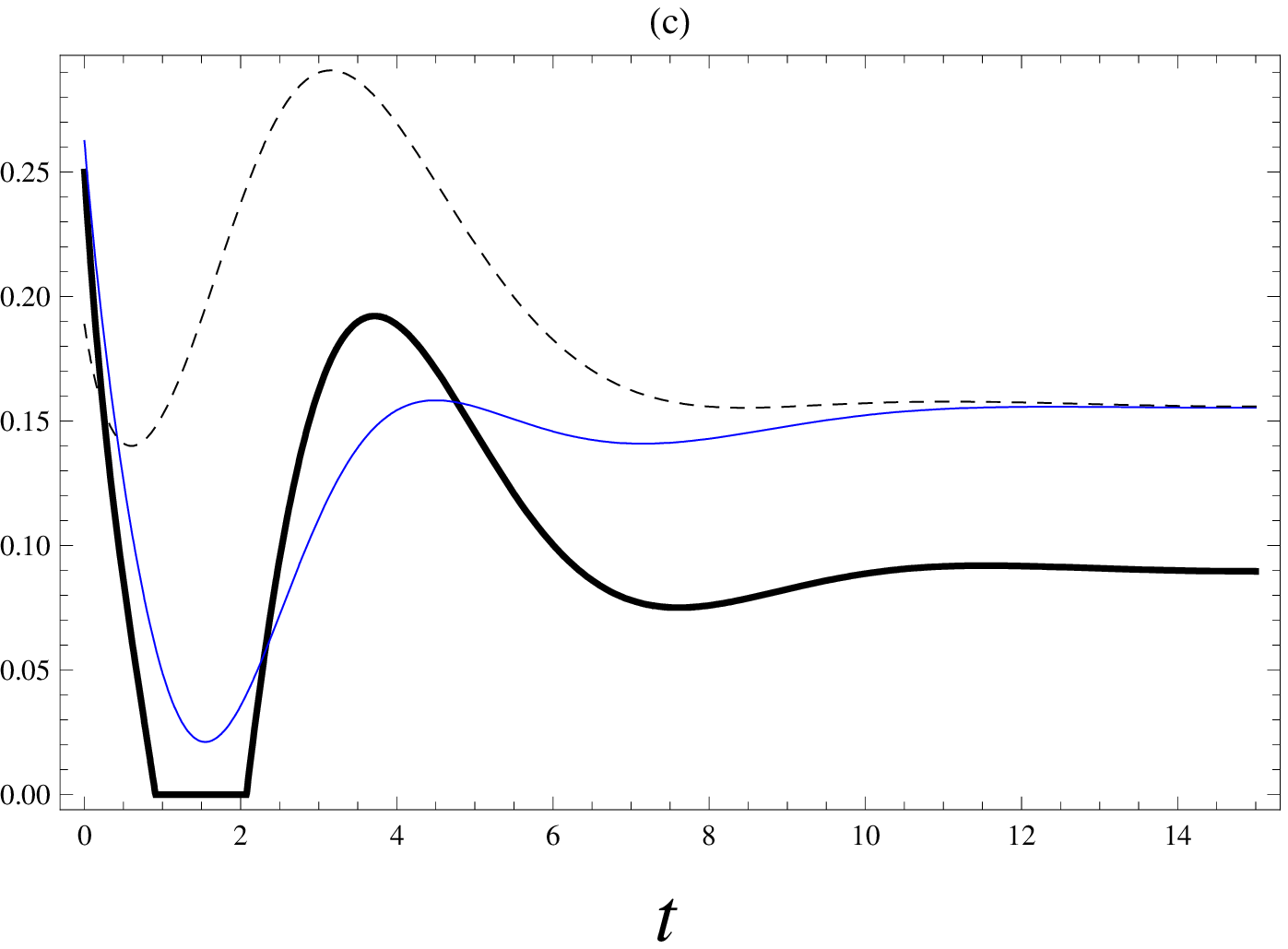}
\includegraphics[width=0.5\textwidth]{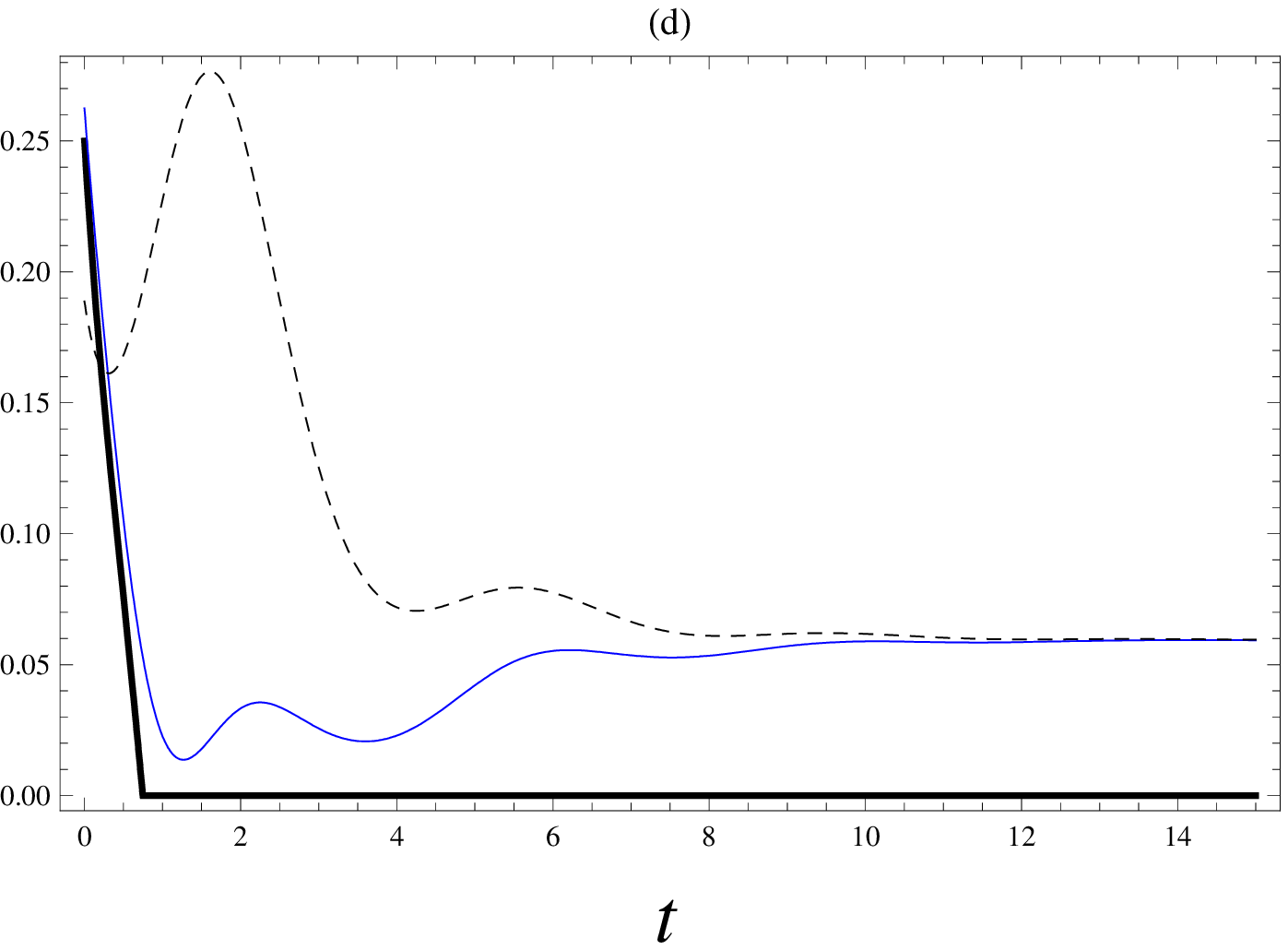}
\caption{Dynamics of concurrence (thick solid line), quantum discord (thin solid line) and classical correlation (dashed line) for the dissipative environment and the other parameters are $p=0.5$, $D=0$ (a) $\Delta=0$, (b) $\Delta=0.2$,  (c) $\Delta=0.4$ and  (d) $\Delta=0.8$.} \label{figure3}
\end{figure}

\begin{figure}[ht]
\includegraphics[width=0.5\textwidth]{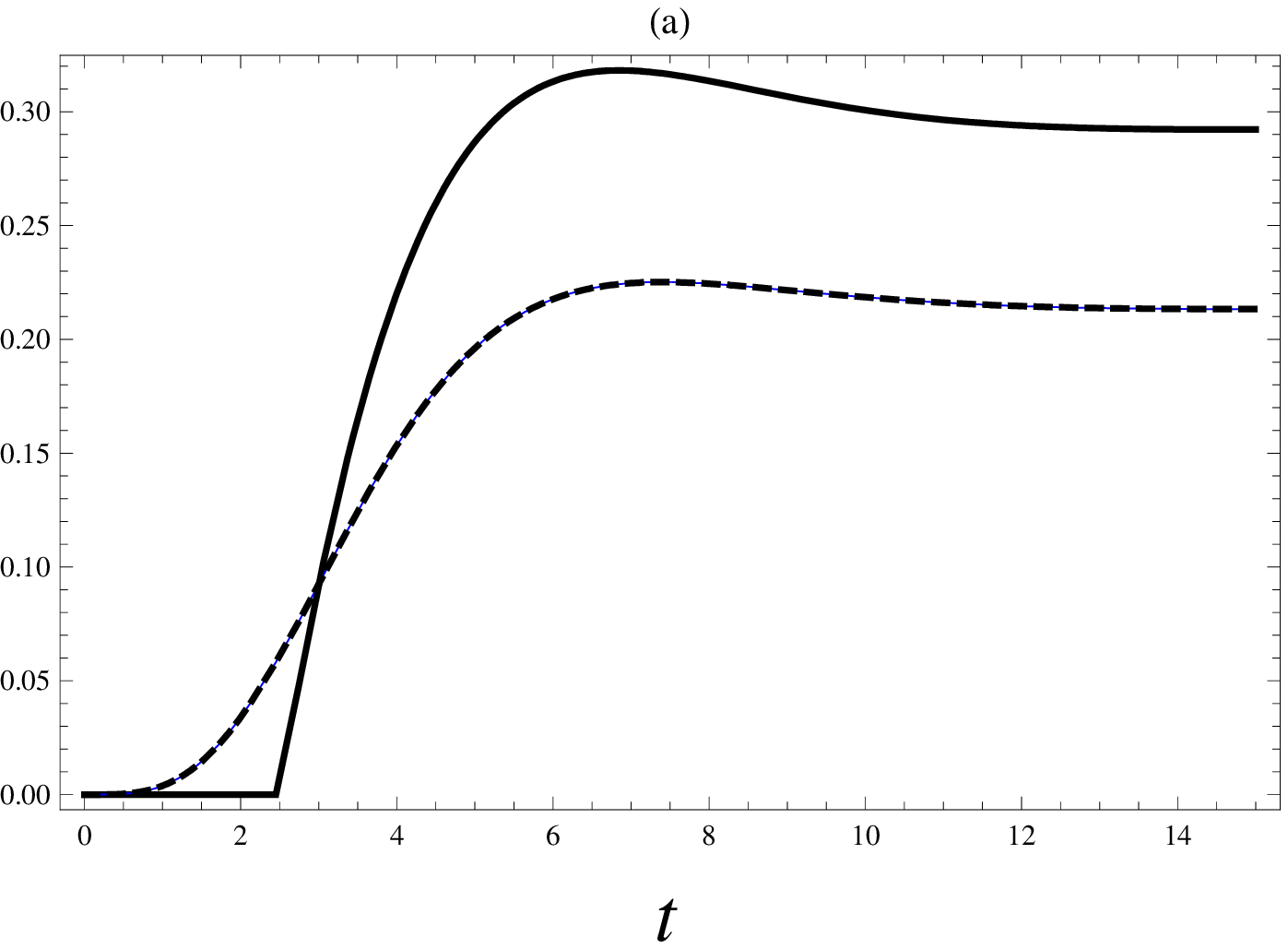}
\includegraphics[width=0.5\textwidth]{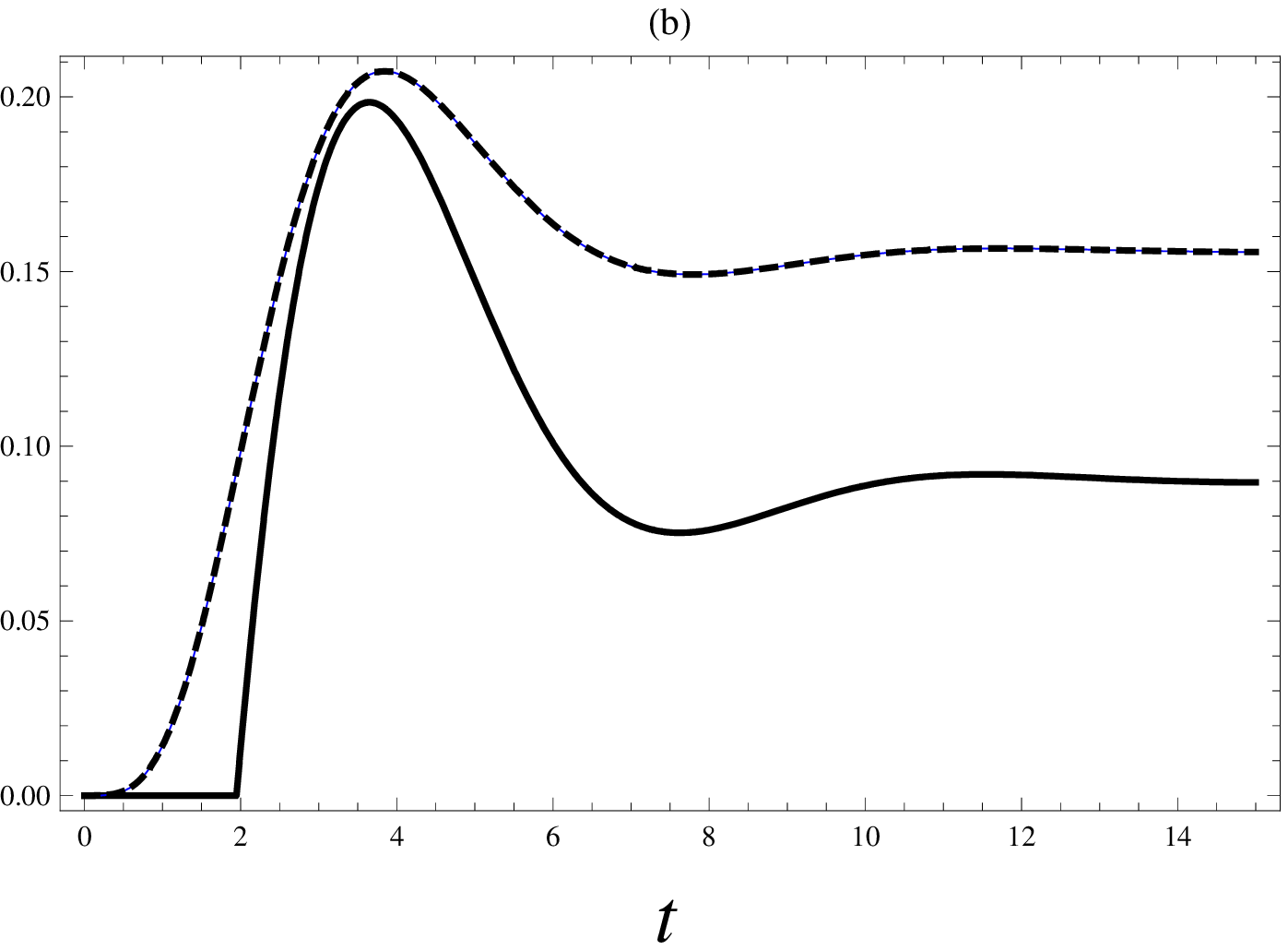}
\includegraphics[width=0.5\textwidth]{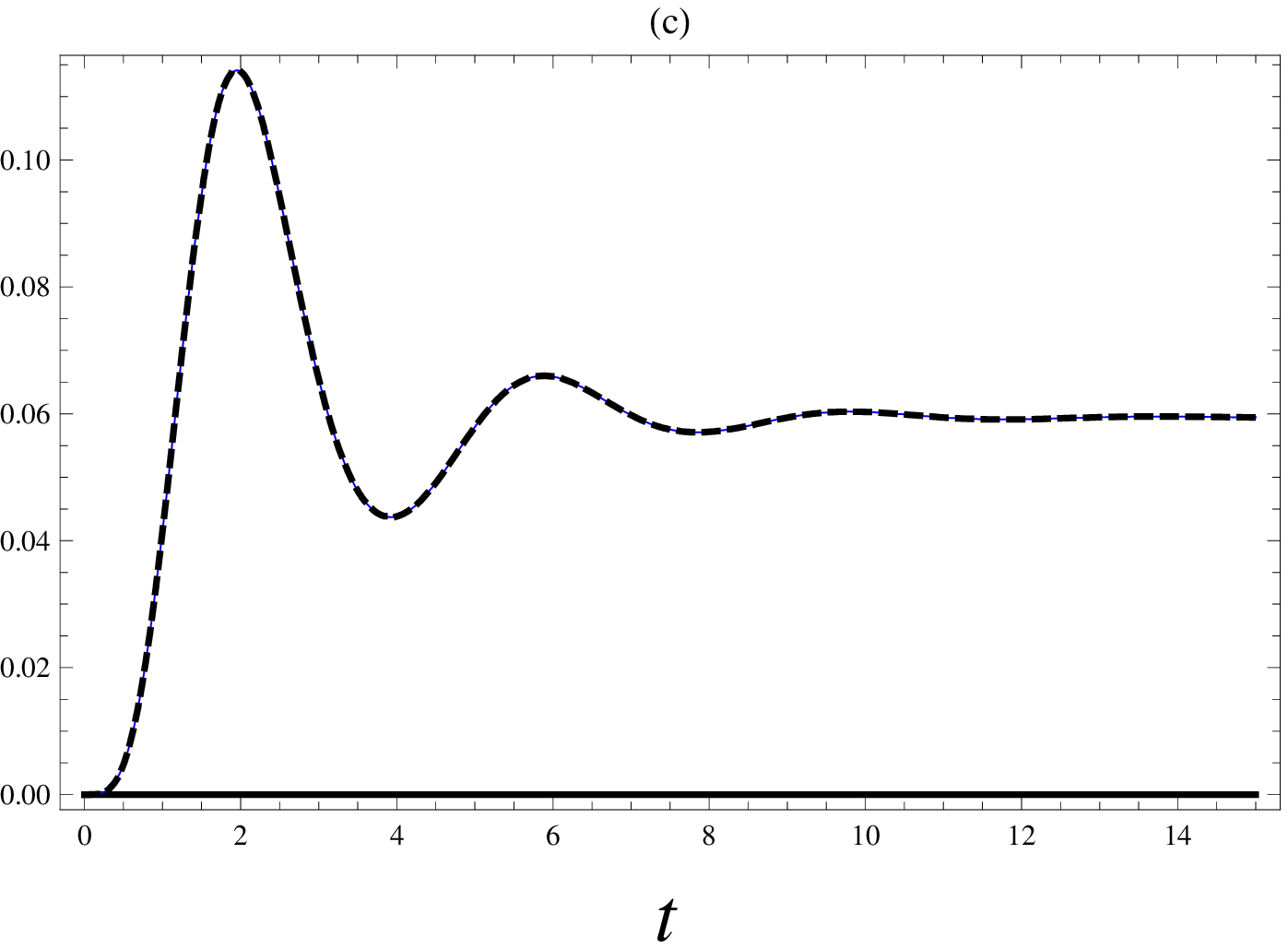}
\caption{Dynamics of concurrence (thick solid line), quantum discord (thin solid line) and classical correlation (dashed line) for the dissipative environment for $p=0$(a) $\Delta=0.2$, (b) $\Delta=0.4$ and (c) $\Delta=0.8$. Note that quantum discord and classical correlation evolve in the same way.} \label{figure4}
\end{figure}

 \begin{figure}[ht]
\includegraphics[width=0.5\textwidth]{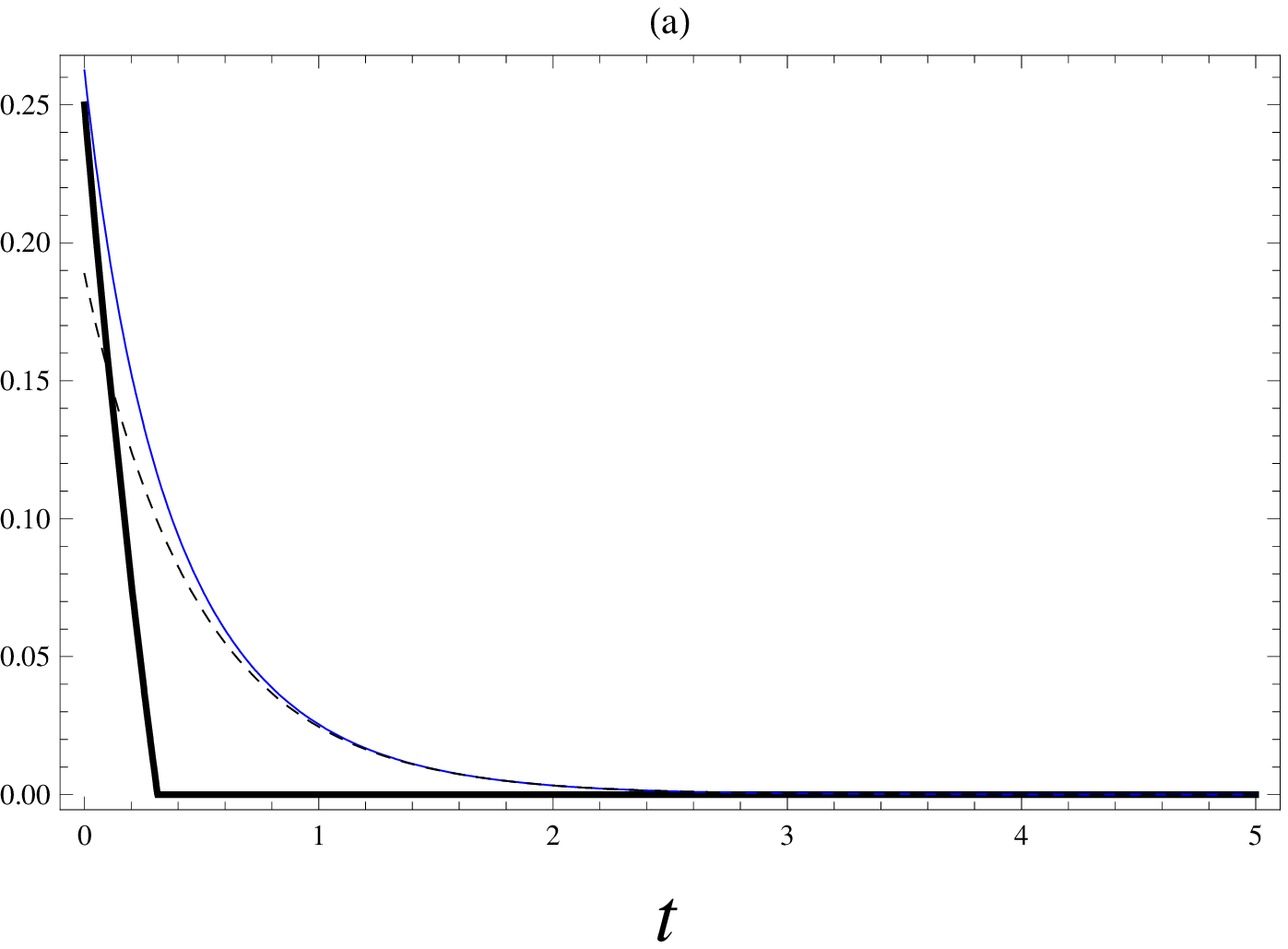}
\includegraphics[width=0.5\textwidth]{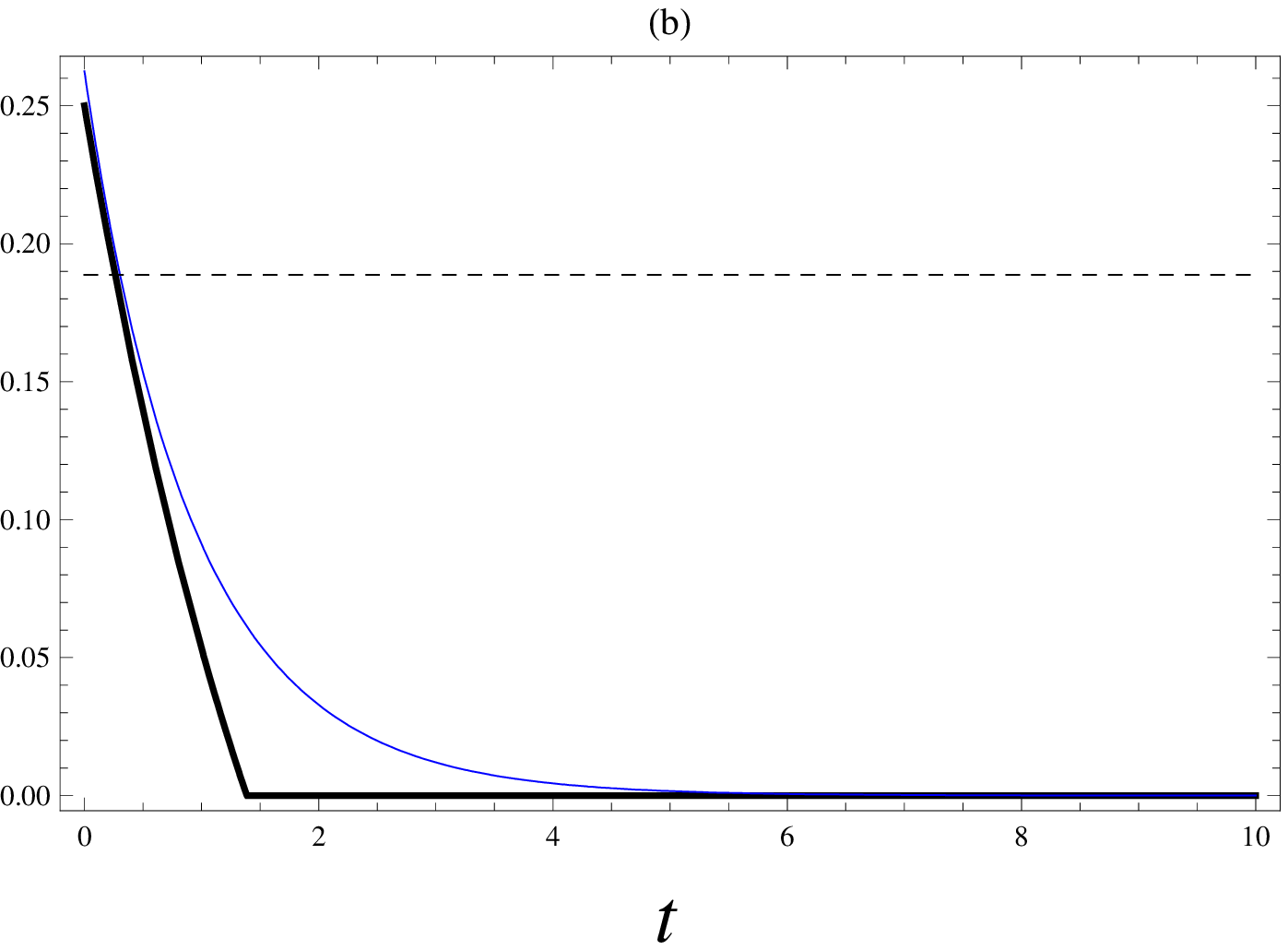}
\caption{Concurrence (thick solid line), quantum discord (thin solid line) and classical correlation (dashed line) as a function of time for $p=0.5$, $D=0$ and (a) noisy environment, (b) dephasing environment.} \label{figure5}
\end{figure}


\begin{thebibliography}{99}
\bibitem{label1} Einstein, A., Podolsky, B., Rozen, N.: Can quantum-mechanical description of physical reality be considered complete?  Phys. Rev. {\bf 47}, 777 (1935) 
\bibitem{label2} Schr\"{o}dinger, E.: Die gegenw\"{a}rtige situation in der quantenmechanik. Die Naturwissenschaften. {\bf 23}, 807 (1935) 
\bibitem{label3} Bell, J.S.: On the Einstein Podolsky Rosen paradox.  Physics. {\bf 1}, 195 (1964)
\bibitem{label4} Nielsen, M.A., Chuang, I.L.  Quantum computation and quantum information. Cambridge: University Press, in UK (2000)

\bibitem{label5} Bennett,  C.H., Brassard, G., Crepeau, C., Jozsa, R, Peres, A., Wootters, W.K.: Teleporting an unknown quantum state via dual classical and
Einstein-Podolsky-Rosen channels. Phys. Rev. Lett. {\bf 70}, 1895 (1993) 
\bibitem{label6} Bose, S.: Quantum communication through an unmodulated spin chain. Phys. Rev. Lett. {\bf 91}, 207901 (2003)
\bibitem{label7} Ekert, A.K.: Quantum cryptography based on Bell's theorem.  Phys. Rev. Lett. {\bf 67}, 661 (1991)
\bibitem{label8} Yu, T., Eberly, J.H.: Qubit disentanglement and decoherence via dephasing. Phys. Rev. B {\bf 68}, 165322 (2003)
\bibitem{label9} Yu, T.,  Eberly, J.H.: Finite-time disentanglement via spontaneous emission. Phys. Rev. Lett. {\bf 93}, 140404 (2004) 
\bibitem{label10} Ferraro, A., Aolita, L., Cavalcanti, D., Cucchietti, F.M., Acin, A.: Almost all quantum states have nonclassical correlations. Phys. Rev. A {\bf 81}, 052318 (2010)
\bibitem{label11} Girolami, D., Adesso, G.: Interplay between computable measures of entanglement and other quantum correlations. Phys. Rev. A {\bf 84}, 052110 (2011)
\bibitem{label12} Al-Qasimi, A., James, D.F, V.: Comparison of the attempts of quantum discord and quantum entanglement to capture quantum correlations. Phys. Rev. A {\bf 83}, 032101 (2011)
\bibitem{label13} Facchi, P., Florio, G., Pascazio, S., Pepe, F.: Local Hamiltonians for maximally multipartite-entangled states.  Phys. Rev. A {\bf 82}, 042313 (2010)
\bibitem{label14} Jafarpour, M., Pourkarimi, M.R., Akhound, A.: Entanglement sudden death and its suppression in multi-qubit channels, using a magnetic field. IL Nuovo Cimento. B {\bf124}, 269 (2009)
\bibitem{label15} Pourkarimi, M.R., Rahnama, M.: Quantum teleportation under the effect of dissipative environment and hamiltonian XY model. Int. J. Theor. Phys {\bf 53}, 1415 (2014)
\bibitem{label16} Montakhab, A., Asadian, A.: Dynamics of global entanglement under decoherence. Phys. Rev. A {\bf 77}, 062322 (2008)
\bibitem{label17} Montakhab, A., Asadian, A.: Multipartite entanglement and quantum phase transitions in the one-, two-, and
three-dimensional transverse-field Ising model. Phys. Rev. A {\bf 82}, 062313 (2010)
\bibitem{label18} Zeng, H.F., Shao, B., Yang, L.G., Li, J., Zou, J.: Entanglement sudden death induced by the Dzialoshinskii-Moriya interaction. Chin. Phys. B {\bf 18}, 3265 (2009)
\bibitem{label19} Carvalho, A.R.R., Mintert, F., Buchleitner, A.: Decoherence and multipartite entanglement. Phys. Rev. Lett. {\bf 93}, 230501 (2004)
\bibitem{label20}  Shan, C.J., Cheng, W.W., Liu, T.K., Liu, J.B., Wei, H.: Sudden death, birth and stable entanglement in a two-qubit Heisenberg XY
spin chain. Chin. Phys. Lett. {\bf 25}, 3115 (2008) 
\bibitem{label21} Werner, R.F.: Quantum states with Einstein-Podolsky-Rosen correlations admitting a hidden-variable model. Phys. Rev. A  {\bf 40}, 4277 (1989)
\bibitem{label22} Lindblad, G.: On the generators of quantum dynamical semigroups. Math. Phys. {\bf 48}, 119 (1976)
\bibitem{label23} Dzyaloshinsky, I.: A thermodynamic theory of weak ferromagnetism of antiferromagnetics. J. Phys. Chem. Solids. {\bf 4}, 241 (1958)
\bibitem{label24} Moriya, T.: New mechanism of anisotropic superexchange interaction. Phys. Rev. Lett. {\bf 4}, 228 (1960)
\bibitem{label25} Wootters, W.K.: Entanglement of formation of an arbitrary state of two qubits.  Phys. Rev. Lett. {\bf 80}, 2245 (1998)
\bibitem{label26} Ollivier, H., Zurek, W.H.: Quantum discord: a measure of the quantumness of correlations. Phys. Rev. Lett. {\bf 88}, 017901 (2001)
\bibitem{label27} Henderson, L., Vedral, V.: Classical, quantum and total correlations.   J. Phys. A {\bf 34},  6899 (2001)
\bibitem{label28} Luo, S.: Quantum discord for two-qubit systems. Phys. Rev. A {\bf 77}, 042303 (2008)
\bibitem{label29} Mazhar Ali., Rau, A.R.P., Alber, G.: Quantum discord for two-qubit X states. Phys. Rev. A {\bf 81}, 042105 (2010)
\bibitem{label30} Maziero, J., C\'{e}leri, L.C., Serra, R.M., Vedral, V.: Classical and quantum correlations under decoherence. Phys. Rev. A {\bf 80}, 044102 (2009)
\bibitem{label31} Barnett, S.M., Radmore, P.:  Method in theoretical quantum optics. Clarendon Press, Oxford, (1997)
\end{thebibliography}
\end{document}